\documentclass[prb,twocolumn,showpacs,amsmath,amssymb]{revtex4}

\usepackage{graphicx}
\usepackage{dcolumn}
\usepackage{bm}

\begin{document}

\title{Stabilization of vortex-antivortex configurations in mesoscopic \\ superconductors by engineered pinning}

\author{R. Geurts}
\author{M. V. Milo\v{s}evi\'{c}}
\thanks{Presently at: Department of Physics, University Of Bath,
Claverton Down, BA2 7AY Bath, UK}
\author{F. M. Peeters}
\email{francois.peeters@ua.ac.be}

\address{Departement Fysica, Universiteit Antwerpen,
Groenenborgerlaan 171, B-2020 Antwerpen, Belgium}

\date{\today}

\begin{abstract}

Symmetry-induced vortex-antivortex configurations in
superconducting squares and triangles were predicted earlier; yet,
they have not been resolved in experiment up to date. Namely, with
vortex-antivortex states being highly unstable with respect to
defects and temperature fluctuations, it is very unlikely that
samples can be fabricated with the needed quality. Here we show
how these drawbacks can be overcome by {\it strategically placed
nanoholes} in the sample. As a result, (i) the actual shape of the
sample becomes {\it far less important}, (ii) the stability of the
vortex-antivortex configurations in general is {\it substantially
enhanced}, and (iii) states comprising {\it novel
giant-antivortices} (with higher winding numbers) become
energetically favorable in perforated disks. In the analysis, we
stress the potent of strong screening to destabilize the
vortex-antivortex states. In turn, the screening-symmetry
competition favors stabilization of new {\it asymmetric ground
states}, which arise for small values of the effective
Ginzburg-Landau parameter $\kappa^*$.

\end{abstract}


\maketitle

\section{Introduction}
The question of vortex-antivortex (VAV) states in mesoscopic
samples is one of the most intriguing ones in superconductivity
studies of the last decade. Being stabilized in a homogeneous
magnetic field, VAV states are counterintuitive and rather
difficult to explain in usual terms. Nevertheless, by analyzing
the solutions of the linearized Ginzburg-Landau (GL) theory,
Chibotaru {\it et al.} \cite{VAVNature} predicted their stability
in flat superconducting squares as a consequence of the symmetry
of the sample. Namely, in linear theory, the geometry of the
sample boundaries directly translates on the vortex states. As a
consequence, for three fluxons captured by the sample (i.e.
vorticity $L=3$), the vortex state with four vortices and one
antivortex can be energetically more favorable than the triangular
configuration of single vortices. Still, one should bear in mind
that the linear approach of Ref. \cite{VAVNature} is only valid at
the superconducting/normal (S/N) boundary, where $\Psi$ is
extremely small and the non-linear term in the GL theory becomes
negligible. In a further step \cite{VAVJahnTeller,VAVStability},
by taking into account the non-linearity of the first GL equation,
the existence of VAV states was predicted to persist even away
from the S/N phase boundary. The antivortex for $L=3$ in a square
of size $10 \xi_0 \times 10 \xi_0$ was shown to be stable in a
temperature range of about $\Delta T=0.3 T_c$ (with $\xi_0$ being
the coherence length at $T=0$, and $T_c$ the critical
temperature). Up to now, such VAV configurations have not been
realized in experiment. From a theoretical viewpoint, there are
several reasons why these novel vortex states escaped experimental
observation: first, even a tiny defect at the boundary ($<1\%$)
destroys the vortex-antivortex state \cite{VAVEdgeDefect}. Second,
the vortices and the antivortex are all confined in a small area
of typical size less than the coherence length $\xi(T)$. As a
consequence, this proximity results in a strong local suppression
of the order parameter $\Psi$ which makes the separate minima
experimentally undistinguishable - the vortex configuration as a
whole is very similar to a single giant-vortex. And third, imaging
of the magnetic field profile is also shortcoming: the field
generated by the sample is proportional to the supercurrent which
is rather weak in the VAV region (being directly related to
$\Psi$).

Therefore, for future experimental verification of the VAV states,
it is essential: i) to increase the vortex-antivortex distance,
allowing for measurements with realistic spatial resolution; ii)
to obtain a larger contrast in the order parameter, and iii) the
magnetic field, required for e.g. Scanning Tunnelling and
Hall-probe microscopy measurements. However, to observe the 'pure'
VAV nucleation, we should realize this in a \textit{homogeneous}
magnetic field, contrary to the suggestion of Ref.
\cite{VAVMagneticDot} where a magnetic dot was placed on top of
the sample. It is already well-known that VAV states have low
energy in an inhomogeneous magnetic field, such as the stray field
of the magnetic dots \cite{MiloPRL} which may induce antivortices
in its own.

\begin{figure}[b]
\includegraphics[width=140pt]{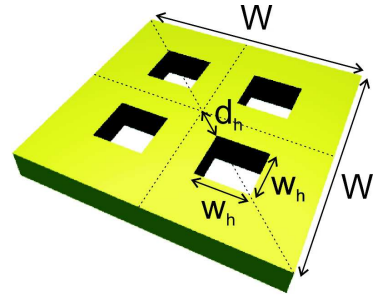}
\caption{\label{fig1} (Color online) Superconducting square with
four nanoholes (i.e. antidots).}
\end{figure}

In our recent Letter \cite{MyPRL}, we introduced the idea of
strategically made perforations/holes/antidots in a square
mesoscopic superconductor (see Fig. \ref{fig1}) which act as
pinning centers for the vortices. The pinning force is supposed to
pull the vortices further from the sample center and thus create a
larger separation between the vortices and the central antivortex.
In what follows, we will elaborate further on this concept, and
demonstrate a number of its advantages. Contrary to Ref.
\cite{MyPRL} where we considered only the square geometry, here we
expand the study to different polygonal shapes of the sample and
geometry of the pinning. In addition, while previous studies
\cite{VAVNature, VAVTriangleLGL, VAV2, VAVJahnTeller} only studied
VAV configurations in extremely thin samples, we consider samples
of finite (though relatively small) thickness, by incorporating
magnetic screening effects in the calculation.

The paper is organized as follows. First the theoretical approach
is formulated (Sec. II), comprising the full non-linear GL theory
and the linearized GL theory (LGL). In Sec. III we present our
results, and determine the optimal pinning parameters to enhance
the stability range of the vortex-antivortex configuration. The
influence of imperfections and defects on the superconducting
state is discussed in Sec. IV, the competition of different
symmetries in the sample is treated (Sec. V), e.g. a disk with a
number of holes. Next we describe the influence of the non-linear
term in GL theory and GL parameter $\kappa$, i.e. screening (Sec.
VI) and some guidelines for a possible experimental observation
(Sec. VII). Finally, we summarize our findings in Sec. VIII.

\section{Theoretical formalism}

\subsection{Ginzburg-Landau theory}

It is already well established that the Ginzburg-Landau (GL)
formalism gives an accurate description of the superconducting state
of low-$T_c$ superconductors, and is well suited to incorporate
boundary effects in the treatment of the mesoscopic superconductors.
The GL theory postulates the following expression for the free
energy of the system (in dimensionless form):

\begin{eqnarray}
\mathcal{F} = \int {\bf dr} \left[ -2|\Psi|^2 + |\Psi|^4 + 2
\left| \left( -i{\bf \nabla} - {\bf A} \right) \Psi \right|^2 \right. \notag \\
 \left. + 2\kappa^{2} \left({\bf B}-{\bf B}_{appl}\right)^2 \right].
\label{GLFreeEn}
\end{eqnarray}
Here all distances are measured in units of the coherence length
$\xi(T) = \xi_0/\sqrt{1-T/T_c}$. The vector potential ${\bf A}$ is
expressed in units of $c\hbar/2e\xi$, the magnetic field in units
of $H_{c2}=\hbar/2e\xi^2$, and the complex order parameter $\Psi$
in units of $\Psi_0=\sqrt{-\alpha/\beta}$, such that $|\Psi|=1$ in
the pure Meissner phase and $|\psi|=0$ in the normal conducting
state (with $\alpha,\beta$ being the GL coefficients
\cite{Tinkham}). $|\Psi|^2$ represents the local Cooper-pair
density, which we will abbreviate as CPD. ${\bf B}_{appl}$ denotes
the applied magnetic field, while ${\bf B}$ stands for the total
magnetic field. $\kappa$ is the ratio between penetration depth
and coherence length, i.e. $\kappa=\lambda/\xi$ and is assumed
temperature independent in this model

To find stable solutions, one has to find the wave function and
the vector potential which minimize the free energy functional.
Two coupled non-linear differential equations, one for the order
parameter and one for the vector potential, can be derived using
variation analysis. We are interested in thin, but
finite-thickness, samples in a perpendicular applied field, where
we may neglect the variation of the magnetic field and order
parameter over the thickness of the sample. Accordingly, we
average the GL equations over the sample thickness \cite{schw,
Tinkham,Prozorov}, and write them as

\begin{eqnarray}
\left( -i{\bf \nabla} - {\bf A} \right)^2 \Psi = \Psi \left( 1 - |\Psi|^2 \right) \label{GLEq1}, \\
-\kappa^{*} \Delta {\bf A} = {\bf j}_s = \Re\left(  \Psi^* (
-i{\bf \nabla} - {\bf A} ) \Psi\right), \label{GLEq2}
\end{eqnarray}
where ${\bf j}_s$ denotes the supercurrent density, and
$\kappa^*=\kappa^2/(d/\xi)$ is the effective Ginzburg-Landau
parameter, the consequence of averaging over sample thickness $d$.
It's worth emphasizing that $\kappa^*$ is temperature dependent
(e.g. through $\xi(T)$ in the denominator). All considered samples
in this paper, treated with the full non-linear GL theory, have the
same thickness $d=\xi_0$. Consequently $\kappa^*$ takes the
temperature dependence $\kappa^*(T) = \kappa^2/\sqrt{1-T}$. To
address properly the influence of $T$, all results obtained by the
full GL theory will be expressed in units of $\xi$, $\psi_0$ and
$H_{c2}$ at zero temperature.

While authors of Refs. \cite{VAVNature, VAVTriangleLGL, VAV2,
VAVJahnTeller} solved only the first GL equation, and restricted
themselves to the limit of $d \rightarrow 0$, or extreme type II
superconductors, we solve numerically {\it both} equations. The
strength of the influence of the second GL equation is governed by
$\kappa^{*}$. When $\kappa^{*} \gg 1$ (thus for $d\ll 1$ and/or
$\kappa\gg 1$), Eq. (\ref{GLEq2}) can indeed be neglected.
However, when $\kappa^{*} \approx 1$, we found it can have a
tremendous influence and provide us with some interesting and
novel behavior of the superconducting state.

After solving Eqs. (\ref{GLEq1}-\ref{GLEq2}), the vorticity $L$ of
the found stable vortex configuration is defined as the sum of the
individual winding numbers of all vortices present in the sample.
The winding number of a vortex is defined as the integer number of
$2\pi$ phase changes of the order parameter when circling around
the vortex core. Consequently, the vorticity of a vortex is
defined as 1, and the vorticity of an antivortex as -1. The total
vorticity of a particular vortex configuration can also be
determined at the sample boundary $\Omega$ by the formula

\begin{equation}
L =-\frac{1}{2\pi}\oint_{\Omega} i \ln( \psi/|\psi| ).
\label{vorticity}
\end{equation}

The vortex state of small mesoscopic superconductors is strongly
influenced by the imposed topological confinement. The shape of
the sample boundaries is introduced in our calculation through the
Neumann boundary condition, which sets the supercurrent
perpendicular to the boundary equal to zero:

\begin{equation}
{\bf n} \cdot \left. \left( -i{\bf \nabla} - {\bf A}\right)
\right|_{\text{boundary}} = 0,
\end{equation}
where ${\bf n}$ is the normal unit vector on the surface.
In Refs. \cite{VAVNature, VAVTriangleLGL, LeuvenGauge} a unitary
transformation was used in combination with a certain gauge for
the vector potential such that for the linear GL equation ${\bf
A}=0$ on the sample boundary. This approach is restricted to
simple geometries (i.e. polygons) and cannot be applied to our
multiple connected samples that contain holes.

Our numerical approach is based on a finite difference
Gauss-Seidel relaxation for the time dependent GL equations. The
numerical technique we applied is the one introduced in Ref.
\cite{Kato} with convenient iterative expansion of the non-linear
term explained in Ref. \cite{schw}, to speed up convergence. We
solve the GL equations on a uniform Cartesian grid typically with
$128 \times 128$ points.

\subsection{Linearized approach}

Since the VAV state is a consequence of the symmetry of the sample
boundary and because of the most efficient reflection of this
symmetry into the solutions of the linearized Ginzburg-Landau
(LGL) theory, the LGL formalism is the ideal instrument to study
the influence of the geometry on the VAV state. From a numerical
point of view, the LGL approach is far less demanding, and can
still provide us with the minimal requirements for the realization
of the VAV states.

The linear theory is \emph{exactly valid only on the S/N
boundary}, i.e. when the $|\Psi|^3$ term can be neglected. Due to
the weak superconducting order parameter, the magnetic field
equals the applied one and the second GL equation can be
discarded. At the same time, Eq. (\ref{GLEq1}) becomes

\begin{equation}
(i {\bf \nabla} + {\bf A})^2 \psi = \alpha \psi, \label{LGLEq}
\end{equation}
where $\alpha$ now is equivalent to an eigenvalue. An important
property of Eq. (\ref{LGLEq}) is that its solution is independent
of the size of the sample, as long as the applied flux is held
constant, i.e. the LGL method is a scalable theory.

To be able to compare the solutions of the LGL obtained for
different geometries, one has to make sure that they have the same
normalization. We use the following normalization of the wave
function, $\frac{1}{V}\int_V dV |\Psi|^2$ = 1, or equivalently
$\langle|\Psi|^2\rangle=1$, where the integration is performed
over the sample volume $V$. This matches the uniform solution of
$|\Psi|=1$ in the whole sample in the absence of a magnetic field.

Our aim is to use the LGL model deeper inside the superconducting
state, even though it is strictly not valid, simply as a limiting
case of the full GL theory. For that purpose we still derive the
generated magnetic field profile using Eq. (\ref{GLEq2}). Although
the LGL theory is not able to determine an absolute order of
magnitude of the field (because $\Psi\approx 0$), it gives the
correct magnetic field up to a multiplication constant, and thus
allows for comparison of field profiles for different sample
geometries.

We compute the vector potential ${\bf A}$ from the calculated
distribution of supercurrents, for taken $\kappa^*=1$. For this
purpose, we solve two independent Poisson equations (for $A_x$ and
$A_y$) using the Fast Fourier Transform. Magnetic field is then
obtained as ${\bf B} = {\bf \nabla} \times {\bf A}$. The relative
magnetic field profile found in this way turns out to be rather
accurate, even when going deeper in the superconducting state
(e.g. by decreasing temperature). When compared with a calculation
based on the full non-linear GL theory, the main effect of the
second equation is to enhance the magnetic field generated by the
sample.

As an advantage, to solve Eq. (\ref{LGLEq}) is computationally not
very demanding, and we use the numerical package COMSOL (formerly
known as Femlab) for this purpose. This software package uses the
finite element method to solve differential equations. Its
profound precision (in solving linear equations) allows us to
significantly increase the grid resolution, for instance, in the
neighborhood of the VAV complex. The use of triangular finite
elements proved as a more accurate treatment of the sample's
geometry than we could achieve with a rectangular grid, for an
arbitrary sample geometry.

Therefore, our linear approach is well suited for comparison of the
VAV state in different samples. We will use it to determine the
optimal hole parameters and to study the influence of imperfections
and defects. We will also treat the competition of different pinning
symmetries and samples with different geometries, like e.g.
perforated disks, in search for novel VAV states.

\section{Engineering of the artificial pinning sites}

\begin{figure}[t]
\includegraphics[width=240pt]{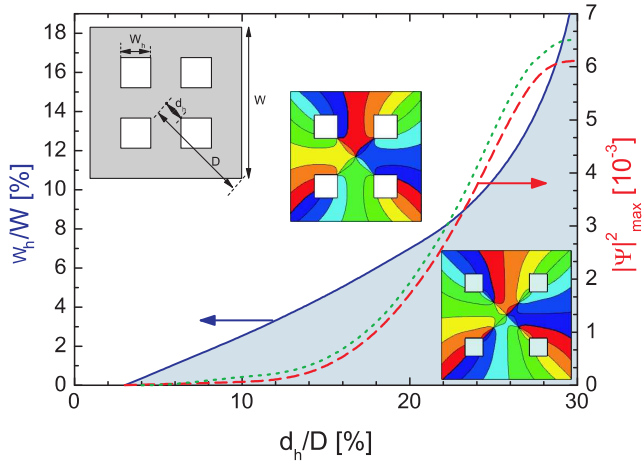}
\caption{\label{OptHolesSizePosSquare} (Color online) Blue solid
curve: The minimum size $W_h$ of the hole for the vortices to be
fully captured by the holes at distance $d_h$. Above the blue
line, vortices are contained in the holes. Below the blue line,
vortices are outside and between the holes. Red dashed curve: The
maximum CPD between the vortex and the antivortex when the holes
are minimum sized. Green dotted curve: Dimensionless quantity
$\Delta B_{vav}$, expressing the strength of the magnetic field
difference measured in the vortex and in the antivortex. Insets
are contour plots of the phase of the wave function (blue/red
corresponds to 0/2$\pi$) for $d_h/D=28\%$ and $w_h/W=18\%$ (upper)
and $d_h/D=36\%$ and $w_h/W=14\%$ (lower). The applied flux
$\phi=5.5\phi_0.$}
\end{figure}

\begin{figure}[t]
\includegraphics[width=240pt]{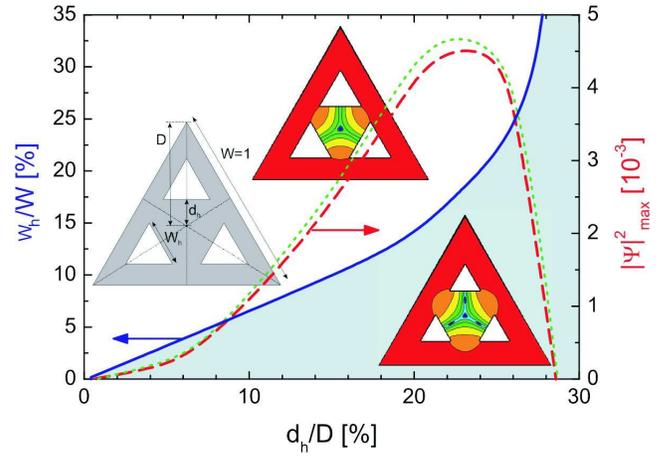}
\caption{\label{OptHolesSizePosTriangle} (Color online) Blue solid
curve: The minimum size $W_h$ of the hole for the vortices to be
fully captured by the holes at distance $d_h$. Above the blue
line, vortices are contained in the holes. Below the blue line,
vortices are outside and between the holes. Red dashed curve: The
maximum CPD between the vortex and the anti vortex when the holes
are minimum sized. Green dotted curve: Dimensionless quantity
$\Delta B_{vav}$, expressing the strength of the magnetic field
difference measured in the vortex and in the antivortex. Insets
are contour plots of the CPD of the wave function (blue/red
corresponds to low/high density - logarithmic color scale) for
$d_h/D=22\%$ and $w_h/W=24\%$ (upper) and $d_h/D=24\%$ and
$w_h/W=18\%$ (lower). The applied flux $\phi=3.5\phi_0.$}
\end{figure}

In what follows, we will study the optimal parameters of the
artificial pinning centra, i.e. size and position of the holes
introduced in the mesoscopic sample, using the linear GL theory.
Later on, the influence of the non-linearity of the GL equations
and the inclusion of the second GL equation will be systematically
investigated.

Our aim is to enhance the parameters that characterize the VAV
state. These are: (i) the maximal Cooper-pair density (CPD)
between the V and AV ($|\psi|_{max}^{2}$) because of the need of
CPD imaging techniques (e.g. STM) for a sufficient contrast in the
local density of states, (ii) the distance between V and AV
($d_{vav}$) which has to be larger than the spatial resolution of
the measurements, and (iii) the magnetic field difference between
V and AV ($\Delta B_{vav}$). All the latter properties for a given
VAV state are defined for the VAV pair with smallest separation
$d_{vav}$.

The influence of the holes on the vortices is determined by their
position and their size. How can we enhance e.g. the
vortex-antivortex separation by manipulating these parameters?
Holes are known to be pinning sites, i.e. they attract vortices.
By placing four holes, we can pull the vortices, which surround
the antivortex, away from the antivortex. The larger the hole, the
stronger the attraction. The forces on the antivortex are exactly
cancelled because of the symmetric position of the holes. However,
the vortices also experience inward forces, i.e. they are
attracted by the antivortex and additionally the Meissner current
also compresses the vortices to the inside of the sample. The
vortices eventually will find an equilibrium position, in which
the inward and outward forces cancel exactly. This position
depends of the distance of the hole to the center and of the size
of this hole.

To understand the influence of the hole parameters (size and
position) we introduce a criterium which we can use as a guide to
determine the optimal hole parameters. This criterium is: What is
the minimum hole size for a given hole position, such that the
vortex is still captured by the hole? It seems that when holes are
made larger, they attract the vortices more strongly and
eventually the vortex will be captured by the hole, when the holes
are placed not too close to the boundary of the sample. We show
our results in Fig. \ref{OptHolesSizePosSquare} and Fig.
\ref{OptHolesSizePosTriangle} for the square and the triangle
geometry, respectively. The insets show the setup and the
definition of the hole parameters. The characteristic variables
$d_h$($=d_{vav}$), $|\psi|_{max}^{2}$ and $\Delta B_{vav}$ are
also shown in Figs. \ref{OptHolesSizePosSquare} and
\ref{OptHolesSizePosTriangle}.

We note that the curves of $|\psi|_{max}^{2}$ and $\Delta B_{vav}$
are isomorphous. We expect this, because $|\psi|_{max}^{2} \propto
|\psi|^2 \propto j_s \propto \Delta B \propto \Delta B_{vav}$. The
maximum $|\psi|_{max}^{2}$ and, as a consequence, $\Delta B_{vav}$
in a triangle appear to be 40 \% less than for the square.

From these figures, we can conclude that, for both the square and
the triangle shaped samples, moving the holes farther from the
center implies enlarging them, in order to keep the vortices
trapped by the holes.
From a certain distance $d_h$, the needed hole size $w_h$
diverges, i.e. from a certain distance of the holes, they can
never be made large enough to trap the vortices that surround the
antivortex. For both the square and the triangle, this distance
$d_h/D$ is about 30\%. While for the square increasing the
distance always implies an increase of $|\psi|_{max}^{2}$ and
$\Delta B_{vav}$, this is not true for the triangle: from a
distance $d_h/D \approx 23\%$ these observables start degrading.

However these results apply to a specific shape of the holes, and
in principle it is not fair to compare the perforated square with
the perforated triangle, since the holes have different shape. For
the square, the trapped vortex is located in the corner of the
square hole; for the triangle this vortex is located in the middle
of a side. Thus, as an alternative approach we reinvestigated both
samples, now with circular holes, and noticed that qualitatively
these results do not differ from the ones with the other shaped
holes. Consequently, we conclude that the general conclusions for
square and triangle are independent of the shape of the holes.

\begin{figure}[t]
\includegraphics[width=240pt]{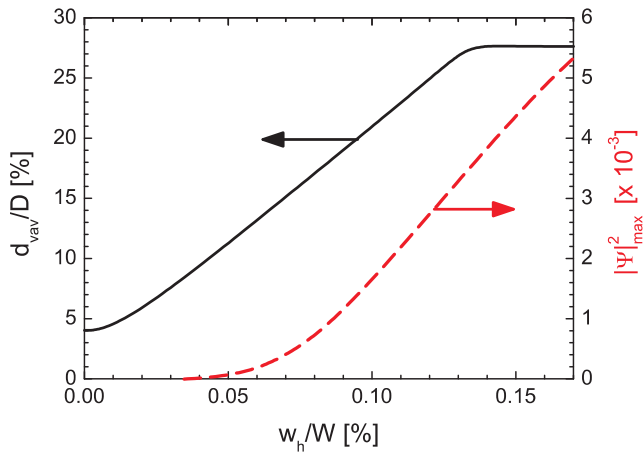}
\caption{\label{HoleSizeEffect} (Color online) VAV distance
($d_{vav}$) and the maximal value of the CPD between the VAV
($|\psi|^{2}_{max} $) versus the hole size $w_h$ for a square with
square holes at position $d_h/D=28\%$. Applied flux $\phi=5
\phi_0$.}
\end{figure}

When, in a realistic situation, one is interested in capturing the
vortices by the holes, one has to chose a somewhat bigger hole
size than the minimal one since small defects will distort the VAV
state and drive a VAV pair closer to each other, hereby the vortex
leaving the hole.

For holes at a distance of $d_h/D=28\%$, the influence of the hole
size is depicted in Fig. \ref{HoleSizeEffect}. However a larger
hole size will of course not increase the VAV distance, it will
increase $|\psi|_{max}^{2}$ which is important for the needed
experimental resolution.

\section{Influence of imperfections/defects}

By introducing nano-engineered holes, we enhanced the
observability of the VAV state. But how about its stability
against defects?

It is generally known that the VAV state in a plain square is very
vulnerable to defects positioned at the edge of the sample
\cite{VAVEdgeDefect}. We will show how this situation dramatically
changes when fourfold symmetric holes are introduced into the
sample. We will consider defects as slight modifications of the
geometry using the linear GL theory. We observe that when the
linear theory predicts that the vortex-antivortex pair
annihilates, it will surely be so in the non-linear theory, which
justifies the use of the LGL theory, in order to determine the
minimal requirements for the stability of the VAV state.

\subsection{A defect at the edge}

As a starting point for comparing the perforated sample with the
plain sample we will study the influence of defects at the edge of
the sample. It is generally known that a small defect at the
boundary will destroy the VAV state in a plain square
\cite{VAVEdgeDefect}. In the present study we restrict ourselves
to defects that are indentations and bulges at the surface of the
sample. We will present here the results for a square for two
different defects positioned at the edge of the sample. The defect
under study is placed at the center of the edge of the square and
is taken to be a square itself with side $w_d$. We considered a
small bulge and a small indentation. It is known that such defects
may influence the penetration and expulsion of vortices as was
demonstrated experimentally by A. K. Geim {\it et. al} in Ref.
\cite{geim}. For theoretical studies we refer to Refs.
\cite{peetersdefect} and \cite{baelusdefect}.

\begin{figure}[t]
\includegraphics[width=240pt]{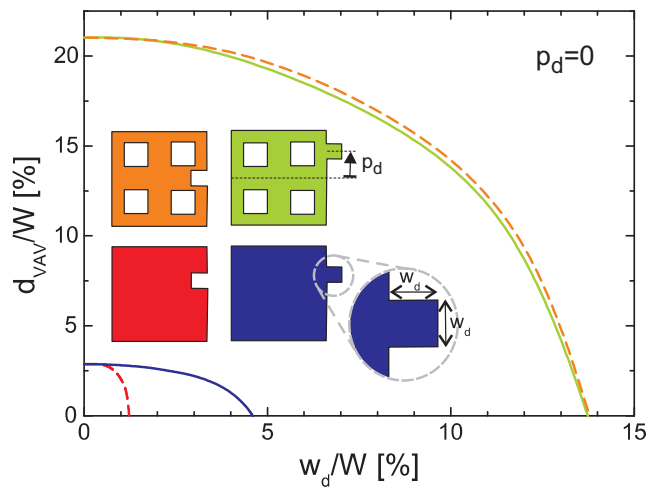}
\caption{\label{EdgeDefectSquare} (Color online) The
vortex-antivortex distance vs defect size for a plain (dark grey)
and a perforated (light grey) square, with a square defect at the
middle of one edge, i.e. $p_d=0$. The dashed curve represents an
indentation, the solid curve a bulge defect. The applied flux for
the plain square $\phi=5.5\phi_0$ and for the perforated square
$\phi=4.0\phi_0$ }
\end{figure}

We focus on the VAV separation $d_{vav}$ to compare the influence
of defect size between the perforated square and the plain square.
When this distance equals zero, the VAV pair annihilates and the
VAV configuration disappears. The results for the plain square are
shown by the two curves at the lower left region of Fig.
\ref{EdgeDefectSquare}. The results for the perforated square are
shown by the two curves at the upper right region. The hole size
and position used for the perforated square are $W_h/W=20\%$ and
$d_h/D=28\%$. For the plain square we applied a flux
$\phi=5.5\phi_0$, for the perforated square $\phi=4.0\phi_0$. We
had to apply different fluxes since the stability range of the L=3
state in both samples do not overlap. (The introduction of holes
shifts the phase diagrams to lower flux.)

For a plain square with a bulge defect, we conclude that the VAV
configuration exists until a size $w_d/W \sim 5\%$. The VAV state
is even more sensitive to an indentation: a size of $\sim 1\%$ is
sufficient for the disappearance of the VAV. However, we found
that the perforated square is much more resistant to defects.
Bulge and indentation defects, up to 14\% (comparable to the size
of a nano-hole), now coexist with a VAV configuration. Bulge and
indentation defects have a similar effect on the VAV stability, in
contrast to the plain square case. We conclude that the 4-fold
symmetrically placed pinning centers are much more efficient in
imposing their symmetry than the outer boundary of the sample. As
a consequence, small defects at the edge have little effect and
only distort the VAV molecule

One may argue that this effect is possibly only since the defect
is located exactly in the center of the edge, therefore imposing
mirror symmetry. Because of this symmetry, the antivortex is
prohibited to chose one of the two vortices of the holes to
annihilate with and it stays on the mirror line.

\begin{figure}[t]
\includegraphics[width=240pt]{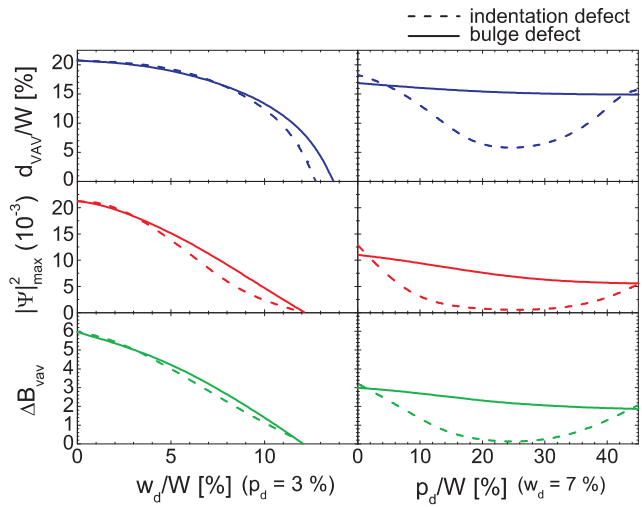}
\caption{\label{DefectSummary} (Color online) The influence of a
defect on the vav distance ($d_{vav}$), CPD between vortex and
antivortex ($|\psi|_{max}^{2}$) and the magnetic field difference
between vortex and antivortex ($\Delta B_{vav}$). The hole
parameters are $w_h/W=20\%$ and $d_h/D=28\%$ and the applied flux
$\phi=4\phi_0.$.}
\end{figure}

To turn off this stabilization effect due to symmetry, we also
investigated defects which are displaced from the middle of the
edge. We studied the influence of the position and size of an edge
defect on the characteristic parameters $d_{vav}$,
$|\psi|_{max}^{2}$ and $\Delta B_{vav}$. The result is shown in
Fig. \ref{DefectSummary} for a perforated square with hole
parameters $w_h/W=20\%$ and $d_h/D=28\%$ under an applied flux of
$4 \phi_0$. On the left side, the influence of the defect size is
depicted for both a bulge and an indentation defect, positioned at
a distance $p_d/W=3\%$ from the edge center. On the right side,
the influence of the position of a defect of size $w_d/W=7\%$ is
shown.

Concerning the influence of the size of the defect, we notice that
bulge and indentation defect act similarly: Increase of the defect
size implies degradation of all three observables
$|\psi|_{max}^{2}$, $d_{vav}$ and $\Delta B_{vav}$. However, for
the position of the defect, we predict different behavior for
bulge and indentation. For a bulge defect, we notice that the
stability of the VAV configuration decreases as the defect moves
further from the center of a side, while for the indentation we
observe a decrease followed by an increase. However, for both
types of defects, the VAV configuration survives best when the
defect is centered at an edge. We also conclude that a bulge
defect is less destructive than an indentation, unless it is
placed near the center of an edge. The antivortex acts like being
attracted to the indentation and repelled by the bulge defect.

\subsection{Other kinds of defects}

The VAV stability against edge defects is strongly improved by the
introduction of the fourfold symmetrically placed holes. However,
in actual experiments these holes themselves can contain defects
or imperfections such as non-uniform sized holes, holes which are
a little displaced with respect to their high symmetry position,
or holes with a slightly different shape.

\begin{figure}[t]
\includegraphics[width=200pt]{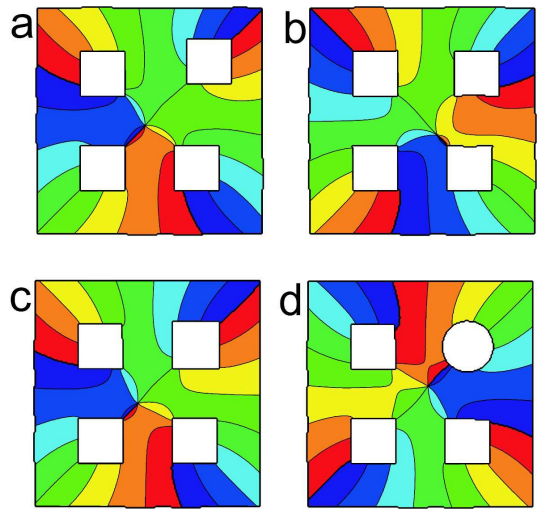}
\caption{\label{SquareOtherDefects} (Color online) Contourplots of
the phase of the order parameter in a perforated square with
several kinds of defects. The hole parameters are: $w_h/W=20\%$,
$d_h/D=22\%$ and applied flux $\phi=4.1\phi_0$. The defects
include: a) a diagonally displaced hole, b) a horizontally
displaced hole, c) a bigger hole and d) a different shaped hole.}
\end{figure}

In Fig. \ref{SquareOtherDefects} four examples are shown. In all
pictures the `normal' holes can be described by the parameters
$w_h/W=20\%$, $d_h/D=22\%$ and we apply a magnetic flux
$\phi=4.1\phi_0$. The plots in Figs. \ref{SquareOtherDefects} (a)
and (b) illustrate the effect of imperfect positioned holes. In
Fig. \ref{SquareOtherDefects} (a), a diagonal displacement of the
upper right hole over a distance $d/W=6\%$ is shown. The VAV
survives such displacements in the range of -4\% to +12\%. For a
horizontal displacement, as illustrated in b), the range is
smaller: from -3.5\% to +3.5\%. The fact that the VAV
configuration is more stable for a diagonal displacement, we
attribute to the existence of mirror symmetry along one diagonal,
forcing the AV on this diagonal.

In Fig.\ref{SquareOtherDefects} (c) the upper right hole has a
somewhat larger size than the other holes. Its sides are $5\%$
larger. This kind of hole size defect does not destroy the VAV
state, when in the range of -8\% to +9\%.

In Fig. \ref{SquareOtherDefects} (d) the upper right hole is
circular. It has the same area as the square holes and is centered
like the square holes. The VAV survives, which illustrates that
it's not the exact shape of the holes which matters, but rather
its area. This implies that non perfect holes, i.e. holes with
defects, will not destroy the VAV state as long as the
imperfection is not too large.

\section{Competing symmetries}

\subsection{Superconducting samples with polygonal pinning}

\begin{figure}[t]
\includegraphics[width=120pt]{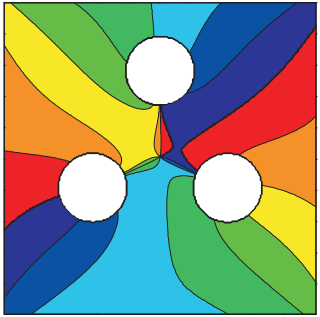}
\caption{\label{Square3HolesAntivortex} (Color online) Contourplot
of the phase of the order parameter for a square sample with three
circular holes. The state with vorticity 2 is shown. The symmetric
positioning of the holes induces an antivortex in the center. }
\end{figure}

Several symmetries compete with each other when the VAV state
nucleates. First we have the vortex-vortex interaction through
which the vortices will try to form the Abrikosov triangle
lattice. Second there is the sample boundary which will impose its
own symmetry, and third there are the pinning centers (i.e. the
holes) which will also try to impose their own symmetry.

For small samples of the order of several coherence lengths, like
the ones we studied up to now, the symmetry of the sample boundary
opposes the Abrokisov lattice. In this subsection we'll point out
that the symmetry of the pinning sites is even more strongly
dominating than the symmetry of the boundary.

For example we show the VAV state with vorticity $L=3-1=2$ in a
square with three holes in Fig. \ref{Square3HolesAntivortex}.
Although the outer edge has a square geometry it is the triangular
arrangement of the circular holes which imposes its symmetry on
the superconducting wave function in the center of the square.
This leads to 3 vortices trapped in the holes and a single
antivortex in the middle of the sample.

\subsection{Superconducting disk with many holes}

The arrangement of holes, instead of the sample outer boundary,
seems to be the geometry element which is much more effective in
imposing its geometry on the wave function. For this reason, we
can as well use circular symmetric disks, perforated by N
symmetrically placed holes, and expect an N-fold symmetric CPD. A
large variety of symmetry induced antivortex configurations can be
created this way. Following this approach we found that it was
possible to even create giant anti-vortices with vorticity $L$ up
to $-7$.

\begin{figure}[t]
\includegraphics[width=140pt]{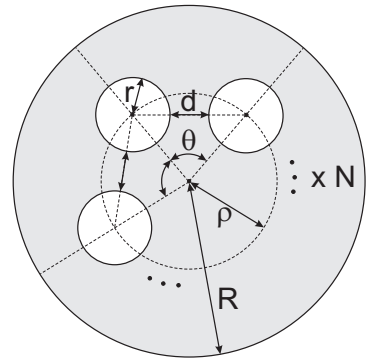}
\caption{\label{CircleNHoles} The parameters $R$, $\rho$, $r$, $d$
and $\theta$, used to characterize the geometry of the disk with N
holes. }
\end{figure}

The parameters which define the geometry are depicted in Fig.
\ref{CircleNHoles}. $d$ is the distance between holes, $\theta$
the angle between holes which is set equal to $2\pi/N$, as to
impose the N-fold symmetry. There is also $\rho$, the distance of
the holes' center to the disks center. $R$ is the radius of the
disk and $r$ is the radius of the holes. Imposing that the
distance between the holes equals the hole's diameter is
equivalent to the condition $r=(1/2)\rho\sin(\theta/2)$.

\begin{figure}[t]
\includegraphics[width=200pt]{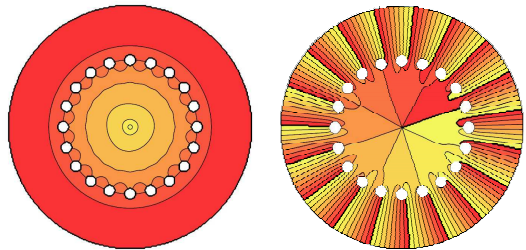}
\caption{\label{Disk20Holes1AV} (Color online) Disk with 20 holes,
in the state with vorticity L=19, consisting of 20 vortices and 1
antivortex in the center. Left: Logarithm of CPD Contour plot (Red
(Dark grey)/Yellow (Light grey) = High/Low density) Right: Phase
Contour plot (Red (Dark grey)/Yellow (Light grey)= 0/2$\pi$) }
\end{figure}

\begin{figure}[t]
\includegraphics[width=200pt]{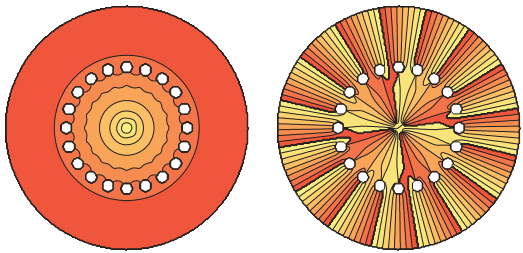}
\caption{\label{Disk20Holes4AV} (Color online) Disk with 20 holes,
showing 4 antivortices in the center. Left: Logarithm of CPD
Contour plot (Red (Dark grey)/Yellow (Light grey) = High/Low
density) Right: Phase Contour plot (Red (Dark grey)/Yellow (Light
grey)= 0/2$\pi$) }
\end{figure}

To illustrate this conjecture, we give some examples. In Fig.
\ref{Disk20Holes1AV} the CPD and the phase plot of a disk with 20
holes is shown. The parameters are $r/R=4.3\%$, $\rho/R=55\%$, so
that the distance between the holes $d=2r$. The applied flux is
$\phi=26\phi_0$. The vorticity $L=19$, is one less the symmetry of
the holes and consequently an additional vortex-antivortex pair is
created. The alternative solution to obey the symmetry would be to
create a $L=19$ giant vortex, but it turns out that the creation
of one VAV pair is energetically more favorable.

There's clearly a competition between the formation of a giant
vortex - the merging of several vortices in one point - and the
formation of a giant antivortex - consisting of the creation of
vortex-antivortex pairs and then merging of all anti-vortices in a
singular point.

For the same disk, with an applied flux of $\phi=21\phi_0$ and
slightly modified parameters $\rho/R=0.5$ and $r/R=4.73\%$ the CPD
and the corresponding phase plot is shown in Fig.
\ref{Disk20Holes4AV}. The phase plot indicates a giant antivortex
in the center with winding number -4. We found that the giant
antivortex is highly unstable with respect to defects. Therefore,
in experiment, it will be very difficult, to observe these giant
antivortices, and most likely they will fall apart in 4 single
antivortices. The reasons for this are: i) a sample with a perfect
symmetry and in the absence of defects will be difficult to
manufacture. In a first stadium, the imperfections will cause the
giant AV to disassociate into separate antivortices and for larger
imperfections it will cause the VAV pairs to annihilate. ii) The
extreme low CPD in the center due to the densely packed
(anti-)vortices which is almost impossible to distinguish between
the separate minima.

\begin{figure}[t]
\includegraphics[width=240pt]{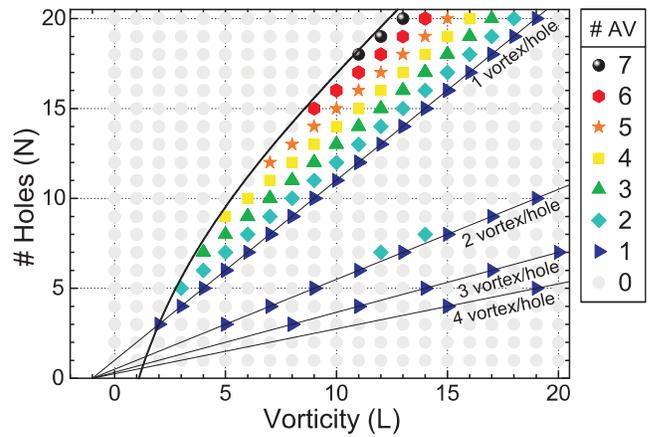}
\caption{\label{NholesAVPD} (Color online) For each vorticity and
for each number of holes N, the number of anti-vortices is shown.
Due to the symmetry the antivortices combine to a giant
antivortex. Left of the thick line no antivortices are observed. }
\end{figure}

To generalize the concept of giant-antivortices and their
appearance in disks with N symmetrically placed holes, we
investigated the relation between the symmetry order N and the
vorticity L. The result is the L-N phase diagram shown in Fig.
\ref{NholesAVPD} where we used the following parameters:
$\rho/R=0.55$, $r=2d$ and $\phi = 1.2 N \phi_0$.

The anti-vorticity of the state is indicated by the different
colors. When $L<N$, and there are no anti-vortices, a giant vortex
of vorticity L is located at the disk center, since this is the
only solution which is able to obey the symmetry. However, because
of the finite grid resolution, these vortices split up in separate
ones, analogous to the fate of the giant antivortex described
before.

For vorticities $L \geq n  N$ (where $n$ is the largest integer
obeying the condition) another property is observed: every hole
pins $n$ vortices, which are not necessarily captured inside the
holes, but clearly belong (i.e. are attached) to the specific
hole.

The L-N phase diagram is not uniquely determined by N. The choice
between giant- or anti-vortices is strongly dependent on the exact
choice of the geometry (i.e. of $r$, $\rho$, $d$) which strongly
affect the free energy, and last but not least of the applied
flux, because of the Meissner current compressing the
(anti-)vortices inward.

\section{Influence of the magnetic screening}

The non-linearity of the GL equations and the coupling to the
magnetic field will now be taken into account. This means that
from now on, we will use the full non-linear GL theory. We will
focus only on the square geometry, but all conclusions can be
extrapolated to other geometries/symmetries. We remind you that
the thickness of the sample throughout this paper is chosen to be
$1 \xi_0$.

The VAV state is now reached by, for instance, increasing the
temperature. In this case the multivortex state will, through a
second order transition, transform into the symmetric VAV state.
In between these two phases, a new state arises: the
\emph{asymmetric} VAV state. This is a stable ground state
configuration, consisting of several vortices and one antivortex,
but they are not positioned symmetrically. The area in a $\phi-T$
phase diagram, where these asymmetric symmetry-induced states are
the ground state configuration, increases when $\kappa$ decreases,
and their existence is therefore a consequence of the second GL
equation.

As a definition for the lower temperature of the asymmetric VAV
region, we take the start of the nucleation of the
vortex-antivortex pair. See inset B in Fig. \ref{KTPD} for an
illustration of the CPD at this point. In our simulations we
noticed a sharp and sudden drop of the CPD in the point where
later the vortex and antivortex are separating. At this point
where the phase change around this pixel is still zero, we define
the beginning of the VAV nucleation when the CPD drops with a
factor
$< 2000 \psi_0/T_c$. We say that a VAV state is symmetric, when $a$,
our measure of asymmetry is smaller than 0.05, which is defined as
$a = \sum\limits_{i} \min\limits_{j\neq i} | {\bf x}_i - \textbf{R}
{\bf x}_j | / \Bigl( \sum\limits_{i} |{\bf x}_i| \Bigr)$. Here,
$\textbf{R}$ is the rotation operator, which rotates a vector over
90 degrees, ${\bf x}_i$ are the positions of (anti-)vortices (when
they're contained in holes, the position is taken to be in the inner
corner pixel of the hole). The origin of the axes is chosen in the
middle of all vortices, i.e. $\sum\limits_{i} {\bf x}_i=0$.

\begin{figure}[t]
\includegraphics[width=240pt]{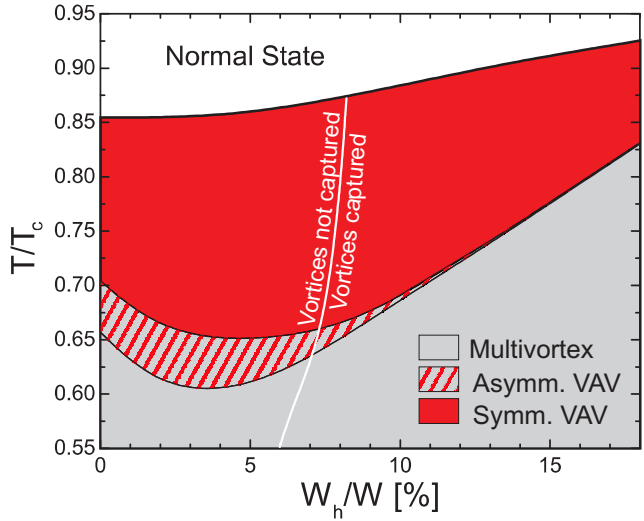}
\caption{\label{HSvsTPD} (Color online) Transition temperatures
$T_{\text{vav}}$, $T_{\text{vav-symm}}$, $T_{\text{S/N}}$  vs. the
size of the hole for a square of sizes $10 \xi_0 \times 10 \xi_0$
and an applied field of $\phi/\phi_0=5.0$. The corners of the
holes are at a distance $d_h/D=25\%$. On the left side of the
white line, the vortices are captured by the holes, on the right
side they are outside the holes. $\kappa=1.5$}
\end{figure}

In the non-linear theory we will now review the role played by the
hole size. In Fig. \ref{HSvsTPD} a phase diagram shows the
dependence of the VAV stability temperature interval of the hole
size. In this figure we used $\kappa=1.5$ and the total applied
flux $\phi$ equals $5.0\phi_0$. The position of the inner corner
of the holes was fixed at $d_h/D=25\%$. Although increasing the
hole size, enhances the characteristic observables
$|\psi|_{max}^{2}$ and $\Delta B_{vav}$ and as well stabilizes
against all kinds of defects, we notice one backdraw of large
holes: the VAV temperature interval shrinks. One can also see that
the VAV temperature interval reaches a maximum size at about
$w_h/W=5\%$ while the temperature range of asymmetric VAV states
remains almost unaltered before this maximum.

\begin{figure}[t]
\includegraphics[width=240pt]{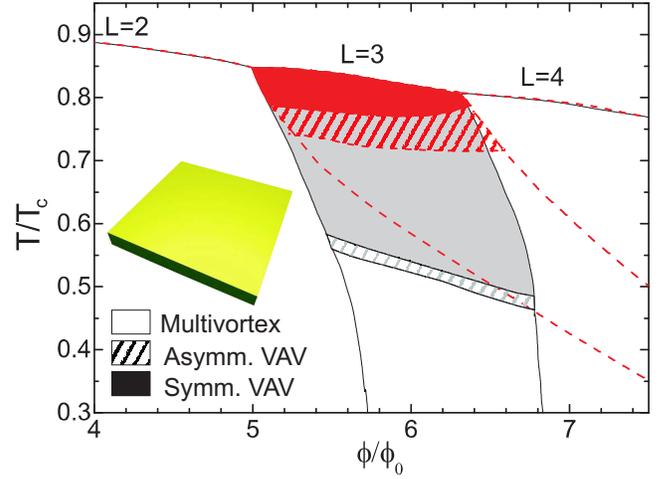}
\caption{\label{PDPlainSquare} (Color online) $\phi-T$ phase
diagram for a plain square, illustrating the ground state regime
of the L=3 state. Grey represents $\kappa=\infty$, red (dark grey)
$\kappa=1$.}
\end{figure}

\begin{figure}[t]
\includegraphics[width=240pt]{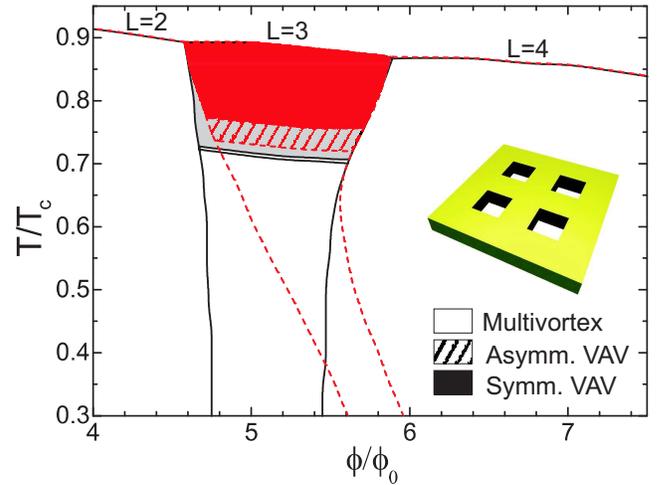}
\caption{\label{PDSquareHoles16CK} (Color online) $\phi-T$ phase
diagram for a square containing four symmetrically placed
nanoholes, illustrating the ground state regime of the L=3 state.
Grey represents $\kappa=\infty$, red (dark grey) $\kappa=1$. Hole
parameters: $w_h/W=12.5\%$ and $d_h/D=25\%$.}
\end{figure}

$\phi-T$ phase diagrams are shown in Fig. \ref{PDPlainSquare} for
a plain square and in Fig. \ref{PDSquareHoles16CK} for a
perforated square, both of size $10 \xi_0 \times 10 \xi_0$. Both
figures contain two superimposed phase diagrams, one for
$\kappa=\infty$ (indicated by grey) and one for $\kappa=1$ (shown
in red (dark grey)). The diagrams illustrate the ground state
region of the L=3 state, so only the neighboring L=2 and L=4
states are taken into account.

Comparing these two phase diagrams with and without holes, one can
clearly see that the introduction of holes of size $w_h/W=12.5\%$
at the position $d_h/D=25\%$ causes the total VAV region to shrink
for a sample with $\kappa=\infty$ but to expand for a sample with
$\kappa=1$. Nonetheless, the region of \emph{asymmetric} VAV
states shrinks for both $\kappa$. Note however that this is not
true for all sizes. See e.g. Fig. \ref{HSvsTPD} where the
asymmetric region initially undergoes a subtle increase.

For both squares a decrease of $\kappa$ leads to: i) a broadening
of the temperature regime with an asymmetric vortex state, and ii)
it causes a shift of the vorticity to higher fields which is large
for low temperatures and smoothly decreases to zero, at the S/N
boundary. The reason for the latter is that the larger currents
($j_s \propto |\psi|^2 \propto 1-T/T_c$) generate a stronger
magnetic field and thus are more effective in expelling and
concentrating the magnetic flux. This shift for decreasing
$\kappa$ can be understood in the limit to type I
superconductivity, where L=0 (i.e. the Meissner state) is the only
stable state. The S/N boundary of both $\kappa$ coincide, since at
the S/N boundary the second equation does not have an influence.

\begin{figure}[t]
\includegraphics[width=240pt]{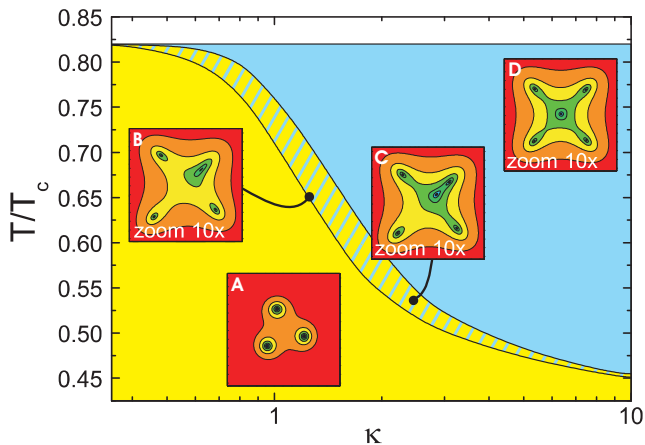}
\caption{\label{KTPD} (Color online) $\kappa$-$T$ phase diagram
showing the regions where symmetric (cyan/dark grey), asymmetric
(shaded) and no antivortex (yellow/light grey) states can be
found. Plain square 10$\xi_0$ $\times$ 10$\xi_0$. Flux
$\phi/\phi_0 = 6.$ Insets show contour plots of the logarithm of
the CPD for the three regions.}
\end{figure}

Fig. \ref{KTPD} depicts the $\kappa$-dependence of the
temperature-interval wherein the (asymmetric) VAV state is stable
in a plain square. It seems that the T-interval of the asymmetric
VAV state grows for decreasing $\kappa$. Also it is shown that the
decrease of $\kappa$ disfavours the vortex-antivortex nucleation,
which is opposite to the findings of Ref. \cite{slava} where this
conclusion was made for a mesoscopic type-I triangle.

For low $\kappa^*$, the supercurrents generate a magnetic field
which contributes to the total magnetic field and destroys its
homogeneity. Obviously, the shape of the magnetic field shows
strong similarities with that of the Cooper pair density, which
could be the reason for the stabilization of asymmetric VAV states
for low $\kappa^*$.

From this point of view we can also interpret the broadening of
the stability region of asymmetric VAV states for small $\kappa$.
The decrease of $\kappa$ acts similarly as having a defect, but
one without preferential spatial direction. It only slows the
nucleation/annihilation of the VAV pair.

Since the vortex-antivortex state already is a state which is
rather unstable and sensitive to all kinds of defects, it is
normal that this subtle equilibrium of the coexistence of vortex
and antivortex disappears and that the VAV pair annihilates.

From a theoretical viewpoint, there are several ways to make the
second order transition from the highly symmetric VAV state to the
multivortex state: by decreasing temperature, the magnetic field
or $\kappa$. In all these three scenarios the same transition
takes place qualitatively: The antivortex
moves towards one of the vortices, they approach and eventually
annihilate. This clearly is a manifestation of spontaneous
symmetry breaking.

\section{How to detect the VAV state?}

\begin{figure}[t]
\includegraphics[width=200pt]{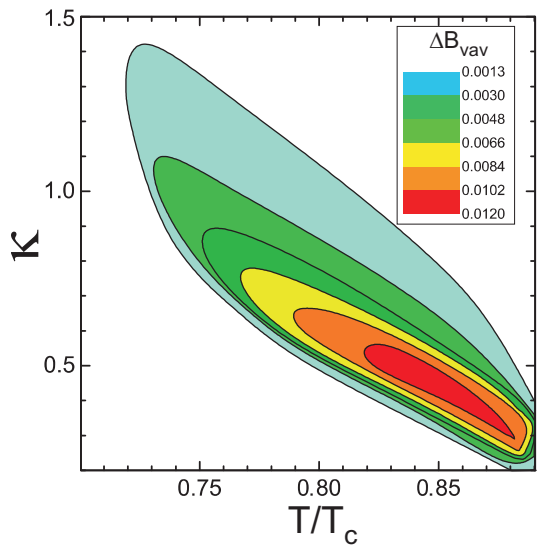}
\caption{\label{DeltaBvsKT} (Color online) Contourplot of the
quantity $\Delta B_{VAV} = (B_{V} - B_{AV}) / B_{appl} $ as
function of $\kappa$ and $T$ for $\phi=5.8$, $T=0.84$ and
$\kappa=0.7$ for a $10\xi_0 \times 10\xi_0$ superconducting square
with holes. Hole parameters: $w_h/W=12.5\%$ and $d_h/D=25\%$.}
\end{figure}

\begin{figure}[t]
\includegraphics[width=240pt]{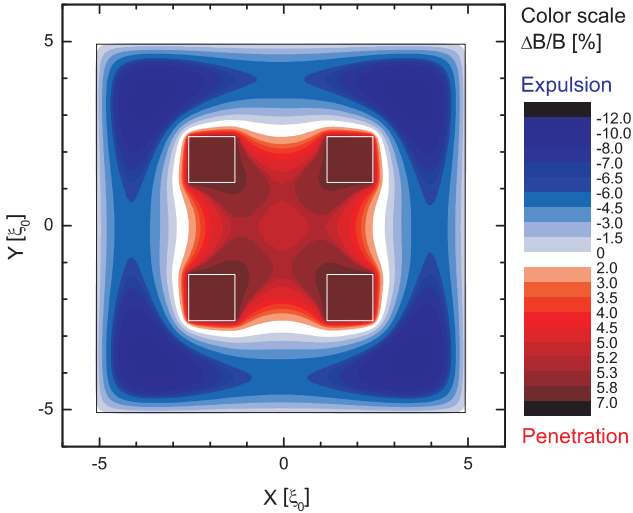}
\caption{\label{DeltaB} (Color online) Contourplot of the quantity
$\Delta B = (B - B_{appl}) / B_{appl} $ for $\phi=5.8$, $T=0.84$
and $\kappa=0.7$ in a $10\xi_0 \times 10\xi_0$ superconducting
square with holes in the L=4-1 state. Hole parameters:
$w_h/W=12.5\%$ and $d_h/D=25\%$.}
\end{figure}

\begin{figure}[t]
\includegraphics[width=240pt]{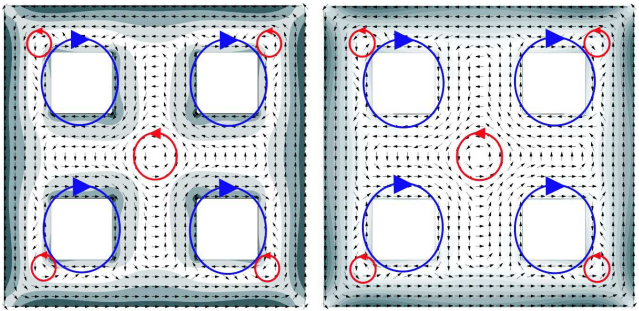}
\caption{\label{CurrentDistributions} (Color online) The
distribution of the supercurrent in a perforated square. The black
arrows indicate the direction of the current, the background
greyscale map depicts the magnitude of the local supercurrent
density (white/black represents low/high). The main
characteristics of the current flow pattern are indicated by the
blue and red circles. Left: $L=4$ state with applied flux
$\phi=7\phi_0$. Right: $L=4-1=3$ (VAV) state with applied flux
$\phi=5.5\phi_0$. }
\end{figure}

\begin{figure}[t]
\includegraphics[width=240pt]{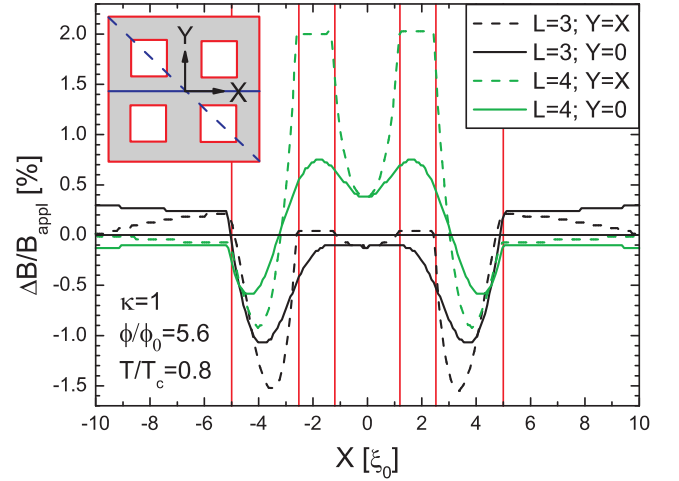}
\caption{\label{L34Field} (Color online) Magnetic field profiles
(diagonal and vertical cut) for the L=4-1=3 and L=4 states in a
perforated square. $\Delta B \equiv B - B_{appl}.$}
\end{figure}

Essentially, we see two ways to proof experimentally the existence
of stable antivortex states: through magnetic field imaging (e.g.
using a Hall-probe) and through Cooper pair density imaging (e.g.
using Scanning Tunneling Microscopy). In this section we will
highlight the advantages and disadventages of both approaches.


An antivortex is characterized by the following properties: i) the
CPD in the center is suppressed to be exactly zero, and ii) its
supercurrents circulate opposite to the one of the vortices.
Unfortunately, the first property also applies to a conventional
vortex, so that it cannot be used to discriminate between vortex
and antivortex. This means that the magnetic field, generated by
the supercurrents, is the only observable parameter able to
distinguish between vortex and antivortex.

In our study of superconductors in a homogeneous field,
antivortices always seemed to appear surrounded by vortices.
However, (multi)vortex currents always generate in the center of
the sample an anti-vortexlike current, in a superconducting cog
wheels motion of the condensate. This current masks the current
profile of a possible antivortex which could be present since it
has the same direction. As a consequence, we can not get evidence
for the existence of an antivortex by looking in a qualitative way
at the magnetic field profile of the sample. As an example, there
is no qualitative difference of the supercurrent and the magnetic
field profile between the L=4 and L=3 VAV state, as is also shown
in Figs. \ref{CurrentDistributions} and \ref{L34Field}. Of course
there is a quantitative difference, which could be exploited.

Nonetheless, one can make use of qualitative differences of both
the CPD and the magnetic field profile, at different temperatures.
Here the challenge is to control the vorticity of the sample and
to assure that it stays constant while measuring and sweeping the
temperature. This way one can either measure the CPD or the
magnetic field profile. At low temperature the magnetic field
profile should clearly exhibit the penetration of the field
through the three vortices residing in three of the holes, while
at a higher temperature the magnetic field should be penetrating
the four holes equally, since a VAV pair is then created, one
vortex occupying the empty hole.

For a more direct observation of the antivortex, we suggest taking
a sample with low effective $\kappa^*$ where a gradual second
order nucleation of the vortex-antivortex pair is found. Then,
using our findings from Sec. VI, subsequent CPD images for slowly
increasing/decreasing temperature or magnetic field should
demonstrate clear evidence for VAV-nucleation/annihilation. In
this scenario, the hole parameters have to be chosen to maximize
the stability region of asymmetric VAV states in the vortex-phase
diagram, following the guidelines from the preceding section.

\section{Conclusions}

The existence of geometry induced antivortices in the presence of
a homogeneous magnetic field has been predicted theoretically
several years ago \cite{VAVNature}. Up to now, there does not
exist any experiment with the observation of it. With numerical
simulations, using both the linear and full non-linear
Ginzburg-Landau theory, we investigated how to engineer
superconducting samples to stabilize and enhance the antivortex.

We propose how to engineer the superconducting sample but without
taking away the conceptual novelty of the nucleation of the
vortex-antivortex pair in a \emph{homogeneous} magnetic field, as
opposed to the idea of placing e.g. a magnetic dot on top of the
sample. We pursued the idea to introduce holes which will act as
pinning centers, and in doing so, pull the vortices away from the
anti-vortex and additionally to provide a strong immunity against
imperfections and defects. First we elaborated on the size and
position of these holes. We determined optimal parameters for the
square and triangle geometry. For instance, in a square geometry,
we managed to enlarge the separation between vortex and antivortex
with a factor 4, compared to the case without holes.

We investigated the influence of several kinds of geometrical
defects on the VAV state. For all the imperfections we studied, we
found that the holes cause a substantial increase of the stability
of the VAV configuration with respect to defects. With the
technology which is nowadays available it is possible to
manufacture samples with the desired precision to proof the
existence of VAV states.

The geometry induced antivortices are known to be a consequence of
the symmetry of the sample. Therefore, we focussed on the
competition of different sources of symmetry in the mesoscopic
superconductor. We concluded that the pinning centers are by far
the most efficient in imposing their symmetry, while the role of
the outer boundary of the sample has a less determining role.
Thinking further on this line, we studied circular disks,
perforated by a number of symmetrically placed holes, and found
giant antivortices up to a vorticity of -7. However, we found that
this spectacular configuration, is highly sensitive to
imperfections.

The effect of the non-linearity of the first GL equations and the
magnetic screening represented by the second GL equation was
critically examined, allowing us to investigate the influence of
temperature and non-zero thickness of the sample on the VAV state.

We constructed phase diagrams for different values of $\kappa$ for
the perforated and the plain square system showing the stability
region of the VAV state in the $\phi-T$ plane. For low value of
$\kappa$ the introduction of holes enlarges the stability region.
We also found the remarkable appearance of asymmetric VAV states,
stable in a wide region of the phase diagram. Such asymmetric
states are counterintuitive since the VAV state is known to be a
consequence of symmetry and yet manifests in an asymmetric way
because of the important non-linearity of the GL equations.
Nevertheless, the existence of asymmetric VAV states can be the
key property for an experiment to proof the existence of VAV
states. We revisited our discussion about the hole size in this
new context and found out that the size of the holes has a large
influence on the stability region of the asymmetric states.

We showed that a small value of $\kappa^*$ disfavors the
vortex-antivortex state in all investigated geometries (square,
perforated square, perforated triangle), indicating that the
findings of Ref. \cite{slava} where a vortex-antivortex state in a
type-I triangle is predicted with large VAV-separation cannot be
correct. We mainly concentrated our discussion on square samples
but we believe that the main conclusions are also valid for
triangle samples.

\section{Acknowledgements}

This work was supported by the Flemish Science Foundation
(FWO-Vl), the Belgian Science Policy, the JSPS/ESF-NES program,
and the ESF-AQDJJ network.


\begin{thebibliography}{0}

\bibitem{VAVNature} L.F. Chibotaru, A. Ceulemans, V. Bruyndoncx, and V.V. Moshchalkov, Nature (London) {\bf 408}, 833 (2000).
\bibitem{VAV2} J. Bon\v{c}a and V.V. Kabanov, Phys. Rev. B {\bf 65}, 012509 (2002).
\bibitem{VAVStability} T. Mertelj and V.V. Kabanov, Phys. Rev. B {\bf 67}, 134527 (2003).
\bibitem{VAVJahnTeller} L.F. Chibotaru, G. Teniers, A. Ceulemans, and V.V. Moshchalkov, Phys. Rev. B {\bf 70}, 094505 (2004).
\bibitem{VAVEdgeDefect} A.S. Mel'nikov, I.M. Nefedov, D.A. Ryzhov, I.A. Shereshevskii, V.M. Vinokur, and P.P. Vysheslavtsev, Phys. Rev. B {\bf 65}, 140503 (2002).
\bibitem{VAVMagneticDot} C. Carballeira, V.V. Moshchalkov, L.F. Chibotaru, and A. Ceulemans, Phys. Rev. Lett. {\bf 95}, 237003 (2005).
\bibitem{MiloPRL} M.V. Milo\v{s}evi\'{c} and F.M. Peeters, Phys. Rev. Lett. {\bf 93}, 267006 (2004); {\it ibid.} Phys. Rev. Lett. {\bf 94}, 227001 (2005).
\bibitem{MyPRL} R. Geurts, M. V. Milo\v{s}evi\'{c}, and F. M. Peeters, Phys. Rev. Lett. {\bf 97}, 137002
(2006).
\bibitem{VAVTriangleLGL} L.F. Chibotaru, A. Ceulemans, V. Bruyndoncx, and V. V.
Moshchalkov,  Phys. Rev. Lett. {\bf 86}, 1323 (2001)
\bibitem{LeuvenGauge} L.F. Chibotaru, A. Ceulemans, G. Teniers, and V.V.
Moshchalkov, Physica C {\bf 369}, 149-157 (2002).
\bibitem{Kato} R. Kato, Y. Enomoto, and S. Maekawa, Phys. Rev. B {\bf 47}, 8016 (1993)
\bibitem{slava} V.R. Misko, V.M. Fomin, J.T. Devreese, and V.V. Moshchalkov, Phys. Rev. Lett. {\bf 90}, 147003 (2003).
\bibitem{schw} V.A. Schweigert and F.M. Peeters, Phys. Rev. B {\bf 57}, 13817
(1998); P.S. Deo, V.A. Schweigert, F.M. Peeters, and A.K. Geim,
Phys. Rev. Lett. {\bf 79}, 4653 (1997).
\bibitem{Prozorov} R. Prozorov, E.B. Sonin, E. Sheriff, A.
Shaulov, and Y. Yeshurun, Phys. Rev. B {\bf 57}, 13845 (1998).
\bibitem{Tinkham} M. Tinkham, \textit{Introduction to Superconductivity} (McGraw Hill, New York,
1975).
\bibitem{golib} G.R. Berdiyorov, B.J. Baelus, M.V. Milo\v{s}evi\'{c}, and F.M. Peeters, Phys. Rev. B {\bf 68}, 174521 (2003).
\bibitem{geim} A.K. Geim, S.V. Dubonos, I.V. Grigorieva, K.S.
Novoselov, F.M. Peeters, and V.A. Schweigert, Nature (London) {\bf
407}, 55 (2000).
\bibitem{peetersdefect} F.M. Peeters, B.J. Baelus, and V.A.
Schweigert, Physica C {\bf 369}, 158 (2002)
\bibitem{baelusdefect} B.J. Baelus, K. Kadowaki, and F.M. Peeters,
Phys. Rev. B {\bf 71}, 024514 (2005).

\end{thebibliography}
\end{document}